\documentclass{kluwer}
\usepackage{psfig}
\begin{document}
\newcommand{\rdm}{{\rm\ rad\ m^{-2}}}
\newcommand{\msuny}{{\rm\ M_{\sun}\ y^{-1}}}

\newcommand{\apj}{ApJ}
\newcommand{\apjl}{ApJL}
\newcommand{\baas}{BAAS}

\begin{article}
\begin{opening}
\title{Linear and Circular Polarization from Sagittarius A* and M81*}

\author{Geoffrey C. \surname{Bower}\email{gbower@astro.berkeley.edu}}
\institute{UC Berkeley Radio Astronomy Laboratory}                               




\runningtitle{Sgr A* and M81*}
\runningauthor{Bower}

\begin{ao}
Radio Astronomy Laboratory \\
601 Campbell Hall \\
Berkeley CA 94720 \\
USA
\end{ao} 


\begin{abstract} 
We describe recent results for Sgr A*, M81* and other low luminosity active
galactic nuclei.  We have  conducted linear  and circular polarimetry over
a frequency range of 1.4 to 230 GHz and detected a variety of phenomena.  The
polarization properties of the 
studied sources are substantially different from higher powered AGN.
In the case of Sgr A*, we are able to eliminate ADAFs and Bondi-Hoyle flows
as possible models based on mm $\lambda$ polarimetry.
Our success with Sgr A* demonstrates that we can learn about the nature of accretion
and outflow in these sources with unprecedented detail.  We may also develop
probes of general relativity in the strong-field limit.
\end{abstract}




\end{opening}
\section{Sagittarius A*}
We have engaged in a lengthy program exploring the linear and
circular polarization properties of Sagittarius A* (Bower et al. 1999a, 
Bower, Falcke \& Backer 1999b, Bower et al. 1999c, Bower et al. 2001, 
Bower et al. 2002a, Bower et al. 2003).  We summarize some
of the key results of that program here.

\subsection{Circular $<<$ Linear at cm $\lambda$}

VLA observations between 1.4 and 15 GHz show that circular polarization
dominates linear polarization at these frequencies.  Linear polarization
is not detected to instrumental limits of 0.1\% while circular polarization
is detected at levels as high as 1\%.  This is substantially different
from what is seen in high luminosity AGN, where linear always
dominates circular polarization (see Homan et al., these proceedings). 
   This property has spawned interest in
models that destroy linear polarization 
and convert linear polarization into circular polarization
through Faraday effects (see Beckert et al. and Ruszkowski et al.,
these proceedings).  These models give us a handle on the
low energy end of the electron spectrum as welll as the magnetic field
geometry.

The fractional circular polarization spectrum is highly variable.
In a flaring state, the spectral index is inverted with a maximum
value of $+1$ between 8.4 and 15 GHz.  This suggests that circular
polarization could be detected at very high frequencies but this
has not been seen yet.

\subsection{Sign of Circular is Constant over 20 years}

We have probed the variability of circular polarization on timescales
ranging from hours to decades.  In one epoch, we see that the circular
polarization varied by $\sim 100\%$ in two hours while the total intensity
changed by $\sim 25\%$.  As noted above, these variations are strongest
at the highest observing frequency.  

Archival VLA observations allow us
to show that the circular polarization at 4.8 GHz did not change
significantly over 20 years.  In these observations and in shorter
term observations, the sign of circular polarization remains negative.
We have no detection at any time of a positive circular polarization
from Sgr A*, although the signal sometimes approaches zero.

This suggests that the circular polarization originates in a region
with a fixed magnetic field orientation.  That this field orientation
is associated with a flaring component is striking and indicates that
it may not have the same character as a shocked region in a higher
luminosity jet.

\subsection{Linear $>>$ Circular at mm $\lambda$}

\begin{figure}[tb] 
\center\mbox{\psfig{figure=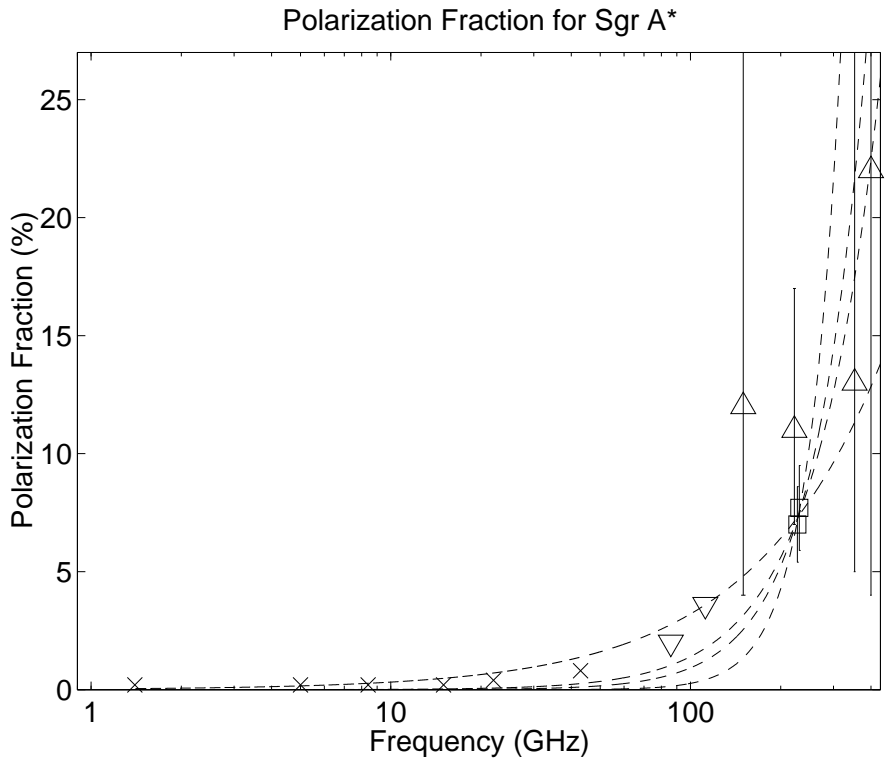,height=4in}}
\caption[]{Linear polarization of Sgr A* as a function of frequency.
Triangles
are the JCMT (Aitken et al. 2000) data.  
Squares are the new BIMA data.  The inverted 
triangles are the 3 mm BIMA upper limits.  
Crosses are VLA upper limits.  All error bars
and upper limits are plotted at 2$\sigma$ levels.  
The dashed lines are power law models
for the polarization fraction with indices of 1.0, 2.0, 2.5 and 4.
}
\label{fig:sgralinpol}
\end{figure}

While there is no circular polarization apparent at millimeter
wavelengths, we have recently detected linear polarization with
the BIMA array at a frequency of 230 GHz (Figure~\ref{fig:sgralinpol}).  
This confirms the JCMT
detection at millimeter and submillimeter wavelengths (Aitken et al. 2000).  
The BIMA observations have an arcsecond beam size, smaller
than the JCMT beam by a factor $>100$.  This allows us to exclude the effects
of polarized dust and unpolarized free-free emission with a great
degree of certainty.

The constant position angle in the upper and lower sideband of the
BIMA observations at 230 GHz
places a strong upper limit to the rotation measure (RM) of the
accretion environment of $2\times 10^6 \rdm$.   An RM this small
excludes a number of models which require large mass accretion rates
onto the black hole.  These include ADAF and Bondi-Hoyle models
which require $\sim 10^{-5} \msuny$.  Thus, the low luminosity of
Sgr A* is due to a low accretion rate rather than to a radiatively
inefficient accretion flow.

We have  a tentative result of RM  $\sim 4\times 10^5 \rdm$
based on the BIMA and JCMT measurements.  However, these measurements
are not fully consistent and re-observation is necessary.  An
actual measurement of the RM will permit us to exclude even lower
accretion rate models.  We will also be able to probe changes in
the accretion environment as a function of time.

\begin{figure}[tb]
\center\mbox{\psfig{figure=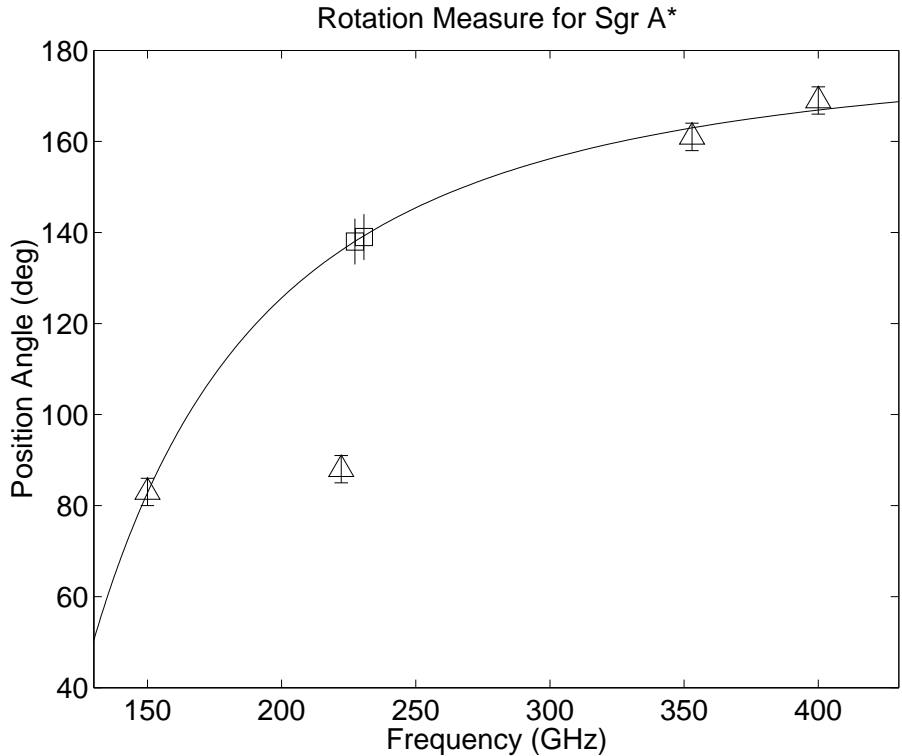,height=4in}}
\caption[]{ Position angle as a function of frequency.  Triangles are
the A00 data.  Squares are the BIMA data.  The solid line is a fit for the RM
excluding the A00 230 GHz result.  The best fit is $-4.3 \pm 0.1 
\times 10^5 \rdm$ with
a zero-wavelength position angle of $181 \pm 2$ degrees.}
\label{fig:rm}
\end{figure}

The linear polarization
spectrum shows a sharp transition from $<1\%$ at 100 GHz to $\sim 10\%$
at 230 GHz.  Bandwidth depolarization is insufficient to account for this
transition while beam  depolarization is marginally adequate.  This requires
a fully turbulent medium with a very small outer scale of turbulence.
An alternative scenario involves an unpolarized low frequency component
and a highly polarized high frequency component with an inverted spectrum.
This is consistent with models that characterize the total intensity spectrum.

The detection of linear polarization in Sgr A* opens up opportunities to 
study the immediate environment of a black hole in substantial detail.  A
number of papers have proposed that general relativistic effects may be
detected (Broderick \& Blandford, these proceedings; Falcke, Melia \& Agol 2000;
Bromley \& Melia 2001).

\newpage
\section{LLAGN}
The polarization properties of Sgr A* are distinct from those of 
high powered AGN.  In particular,
linear polarization dominates circular polarization by typically
an order of magnitude in these sources.  These powerful AGN
are more luminous than Sgr A* by 10 orders of magnitude.  
To determine whether the difference in polarization
is an effect of luminosity, we have studied the linear and
circular polarization of a sample of nearby LLAGN.

\begin{figure}[tb] 
\center\mbox{\psfig{figure=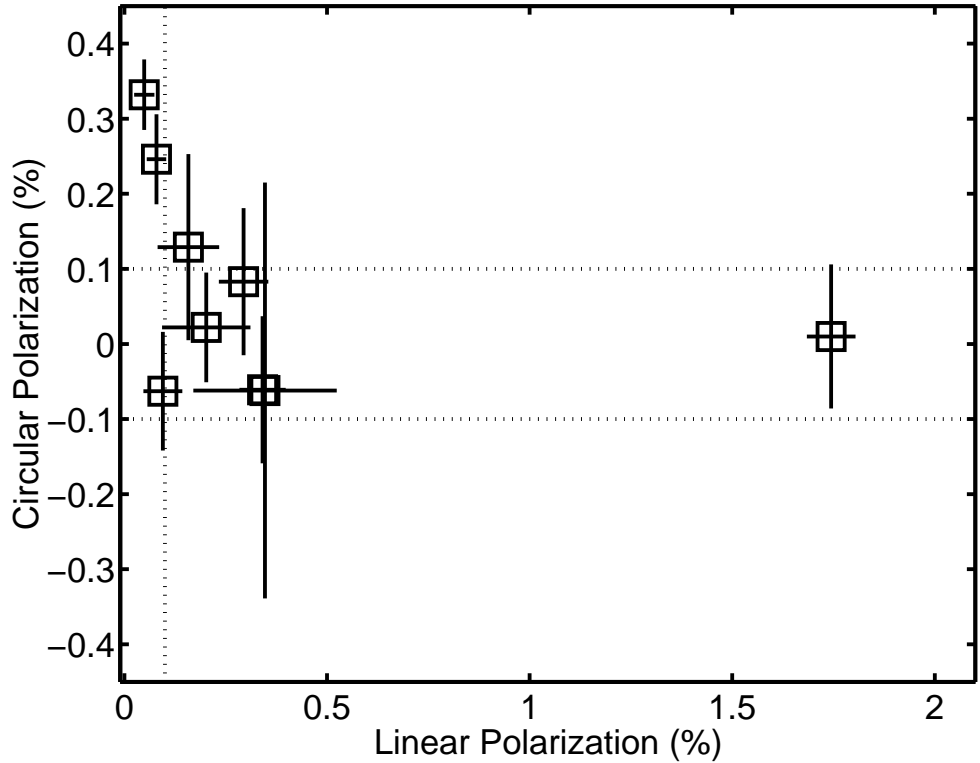,height=4in}}
\caption[]{Linear and circular polarization of low luminosity active
galactic nuclei.  Dotted lines indicate systematic limits of 0.1\%.
Most of the LLAGN show no or very little linear or circular polarization.
}
\label{fig:llagn}
\end{figure}

In Figure~\ref{fig:llagn} we summarize our observations of 9 LLAGN
with the VLA at 8.4 GHz (Bower, Falcke \& Mellon 2002).  
With the exception of M87, all sources showed
weak or no linear polarization.  Linear polarization was only
detected in two sources at a level of a few tenths of a percent.
In NGC 4579 there is a potential detection of RM$=7\times 10^4 \rdm$.
Circular polarization is only detected convincingly in one source
M81*, which we discuss in the next section.

The absence of linear polarization for these sources is 
different from the case of higher luminosity sources and is similar to the case
of Sgr A*.
The result can be readily explained through Faraday depolarization.
An RM$\sim 10^5 \rdm$ can lead to bandwidth depolarization in these
sources at these frequencies.  
An RM$\sim 10^3 \rdm$ can lead to beam depolarization.  We have
demonstrated that both the accretion region and galactic environment
of Sgr A* can readily lead to RMs this large.  

Nevertheless, one cannot exclude the effects of low jet power.  If
low luminosity jets do not exhibit the same degree of field order
as their higher-powered cousins, then they will be weakly polarized.  
They may lack the powerful shocks
that order the magnetic field in high luminosity sources.

We note also that there is a clear trend in spectral index and circular
polarization strength.  Of objects that we have studied, only Sgr A* and M81* 
show circular polarization and only Sgr A* and M81* have inverted spectral
indices.  This supports the hypothesis that circular polarization is
due to an opacity effect in the field.  However, only a small number of
sources are included in this analysis.

\newpage
\section{M81*}
We have explored the linear polarization properties of the LLAGN M81* 
 in more depth (Brunthaler, Bower \& Falcke 2001).  These investigations have shown that the
polarization continues to exhibit similarity with Sgr A* as we observe at
higher frequencies and study the variability properties.  The presence of
a jet in M81* suggests that the polarization properties of
both sources are dominated by a jet.

\begin{figure}[tb] 
\center\mbox{\psfig{figure=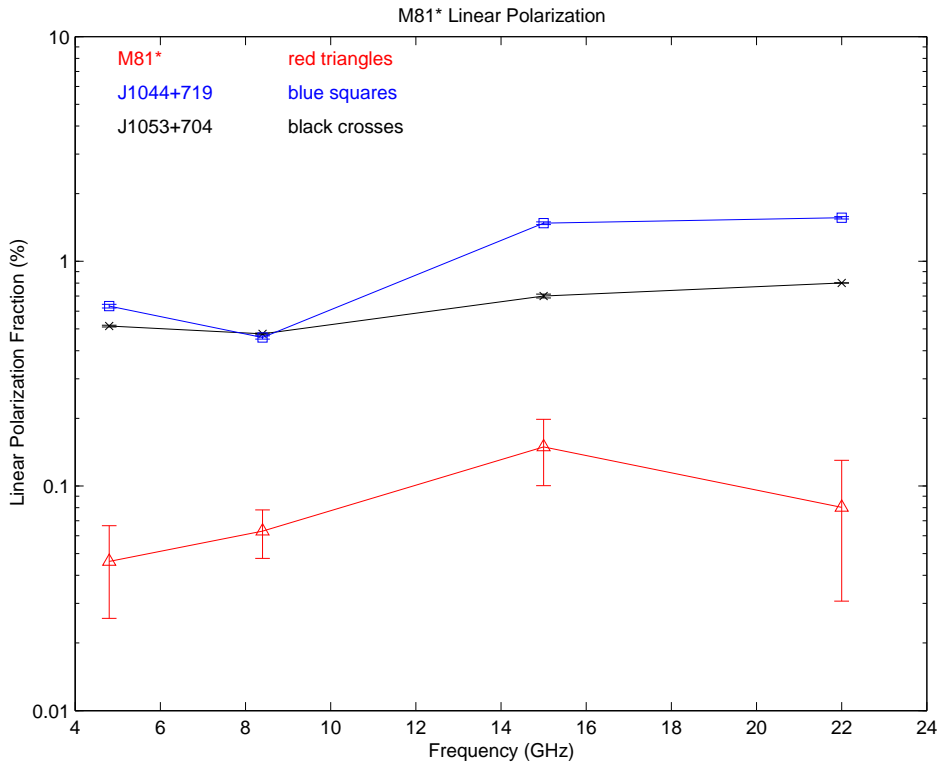,height=4in}}
\caption[]{Linear polarization of M81* and two calibrators as a function
of frequency.  M81* is unpolarized at all frequencies below 22 GHz.
}
\label{fig:m81linpol}
\end{figure}

In Figure~\ref{fig:m81linpol} we show that linear polarization is
absent up to a frequency of 22 GHz.  This implies lower limits
to the RM greater by a factor of $\sim 7$ over those for the 8.4 GHz
survey.  These RM limits are still not surprising given the 
expected particle densities and magnetic field strengths near the
black hole.  

\begin{figure}[tb] 
\center\mbox{\psfig{figure=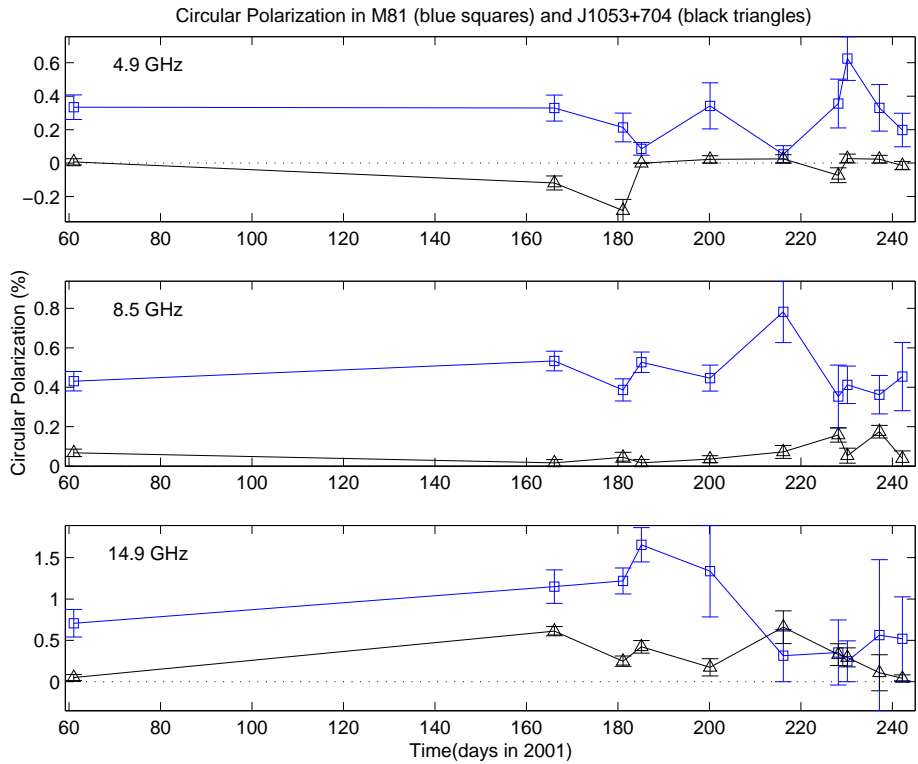,height=4in}}
\caption[]{Circular polarization of M81* and a calibrator as a function
of time at three frequencies.  
}
\label{fig:mc81}
\end{figure}

The circular polarization properties of M81* are also similar
to that of Sgr A* (Figure~\ref{fig:mc81}).   We see that circular polarization
is detected at 4.8, 8.4 and 14.9 GHz with a magnitude as high as 1.5\%.  
The degree of variability increases with frequency.  The lightcurves 
suggest episodic activity that is also characteristic in Sgr A*.  The 
highest point in the circular polarization light curve occurs
$\sim 10$ days after a bright flare in the total intensity.  As the
flare decays, the high frequency circular polarization disappears.

\newpage
\section{Summary}

We have presented here a range of observations of polarization in LLAGN, including
Sgr A* and M81*.  These sources differ markedly from higher luminosity AGN.  
These differences are consistent with small-scale, low-power jets which see the
high density accretion regions and/or galactic HII regions.  Higher
frequency observations may reveal a marked transition in the polarization properties
of LLAGN other than Sgr A* as Faraday effects weaken.  These observations are at
the edge of capability for current millimeter interferometers.  Future arrays
such as CARMA and ALMA will be able to systematically survey a broad sample of LLAGN.
We will be able to 
determine the nature of their accretion environments, including the role of
advection, convection and outflows.  We will be able to explore 
the stability of magnetic field structures,
the presence of black hole spin and general relativistic effects in the vicinity of
the black hole.


\end{article}
\end{document}